\documentclass[conference]{IEEEtran}
\IEEEoverridecommandlockouts
\usepackage[numbers,sort&compress]{natbib}
\usepackage{amsmath,amssymb,amsfonts}
\usepackage{algorithmic}
\usepackage{graphicx}
\usepackage{textcomp}
\usepackage{xcolor}
\usepackage{pgf, tikz, pgfplots}
\usetikzlibrary{positioning}
\usepackage{booktabs}
\usepackage[font=footnotesize]{subcaption}
\usepackage{dblfloatfix}
\usepackage[font=footnotesize]{caption}

\usepackage{siunitx}
\usepackage[nolist]{acronym}
\usepackage{bm}
\usepackage{mathtools}
\usepackage[colorinlistoftodos]{todonotes}
\usetikzlibrary{calc}
\usetikzlibrary{fit}
\usepackage{mleftright}

\newcommand{\RV}[1]{\mathsf{#1}}
\newcommand{\RM}[1]{\bm{\mathsf{#1}}}

\def\j{{\mathrm{j}}}
\def\e{{\mathrm{e}}}
\def\SNR_Y{\mathrm{SNR}_{\RM{Y}}}

\def\BibTeX{{\rm B\kern-.05em{\sc i\kern-.025em b}\kern-.08em
    T\kern-.1667em\lower.7ex\hbox{E}\kern-.125emX}}

\begin{document}
\begin{acronym}[]
    \acro{6G}{sixth generation}
    \acro{AE}{autoencoder}
    \acro{AF}{ambiguity function}
    \acro{AIR}{achievable information rate}
    \acro{AWGN}{additive white Gaussian noise}
    \acro{CA}{cell-averaging}
    \acro{CFAR}{constant false alarm rate}
    \acro{CP}{cyclic prefix}
    \acro{FEC}{forward error correction}
    \acro{FFT}{fast Fourier transform}
    \acro{GMI}{generalized mutual information}
    \acro{IFFT}{inverse fast Fourier transform}
    \acro{ISAC}{integrated sensing and communications}
    \acro{LLR}{log-likelihood ratio}
    \acro{MF}{matched filter}
    \acro{OFDM}{orthogonal frequency division multiplexing}
    \acro{PSK}{phase shift keying}
    \acro{RCS}{radar cross section}
    \acro{QAM}{quadrature amplitude modulation}
    \acro{RV}{random variable}
    \acro{SC}[S\&C]{sensing \& communications}
    \acro{SD}{soft decision}
    \acro{SINR}{signal-to-interference-and-noise ratio}
    \acro{SNR}{signal-to-noise ratio}
    \acro{TOI}{target of interest}
    \acro{FMCW} {Frequency modulated continuous wave}
    \acro{PC-FMCW} {phase coded FMCW}
    \acro{ADC} {analog-to-digital converter}
    \acro{RVPC}{residual video phase compensation}
    \acro{IF}{intermediate frequency}
    \acro{ISNR}{image signal-to-noise-ratio}
    \acro{RDM}{range Doppler matrix}
    \acro{MSE}{mean squared error}
    \acro{CRB}{Cramér-Rao bound}
    \acro{QPSK}{quadrature phase shift keying}
    \acro{64-QAM}{64-quadrature amplitude modulation}
    \acro{DAC}{digital-to-analog-converter}
    \acro{CA-CFAR}{cell averaging constant false alarm rate}
\end{acronym}

\definecolor{kit-gray70}{cmyk}{.6,.6,.6,0}

\definecolor{kit-blue100}{cmyk}{.8,.5.,0,0}
\definecolor{kit-blue70}{cmyk}{.56,.35,0,0}
\definecolor{kit-blue50}{cmyk}{.4,.25,0,0}
\definecolor{kit-blue30}{cmyk}{.24,.15,0,0}
\definecolor{kit-blue15}{cmyk}{.12,.075,0,0}

\definecolor{kit-green100}{cmyk}{1,0,.6,0}
\definecolor{kit-green70}{cmyk}{.7,0,.42,0}
\definecolor{kit-green50}{cmyk}{.5,0,.3,0}
\definecolor{kit-green30}{cmyk}{.3,0,.18,0}
\definecolor{kit-green15}{cmyk}{.15,0,.09,0}

\definecolor{KITgreen}{rgb}{0,.59,.51}

\definecolor{KITpalegreen}{RGB}{130,190,60}
\colorlet{kit-maigreen100}{KITpalegreen}
\colorlet{kit-maigreen70}{KITpalegreen!70}
\colorlet{kit-maigreen50}{KITpalegreen!50}
\colorlet{kit-maigreen30}{KITpalegreen!30}
\colorlet{kit-maigreen15}{KITpalegreen!15}

\definecolor{KITblue}{rgb}{.27,.39,.66}

\definecolor{KITyellow}{rgb}{.98,.89,0}
\definecolor{kit-yellow100}{cmyk}{0,.05,1,0}
\definecolor{kit-yellow70}{cmyk}{0,.035,.7,0}
\definecolor{kit-yellow50}{cmyk}{0,.025,.5,0}
\definecolor{kit-yellow30}{cmyk}{0,.015,.3,0}
\definecolor{kit-yellow15}{cmyk}{0,.0075,.15,0}

\definecolor{KITorange}{rgb}{.87,.60,.10}
\definecolor{kit-orange100}{cmyk}{0,.45,1,0}
\definecolor{kit-orange70}{cmyk}{0,.315,.7,0}
\definecolor{kit-orange50}{cmyk}{0,.225,.5,0}
\definecolor{kit-orange30}{cmyk}{0,.135,.3,0}
\definecolor{kit-orange15}{cmyk}{0,.0675,.15,0}

\definecolor{KITred}{rgb}{.63,.13,.13}
\definecolor{kit-red100}{cmyk}{.25,1,1,0}
\definecolor{kit-red70}{cmyk}{.175,.7,.7,0}
\definecolor{kit-red50}{cmyk}{.125,.5,.5,0}
\definecolor{kit-red30}{cmyk}{.075,.3,.3,0}
\definecolor{kit-red15}{cmyk}{.0375,.15,.15,0}

\definecolor{KITpurple}{RGB}{160,0,120}
\colorlet{kit-purple100}{KITpurple}
\colorlet{kit-purple70}{KITpurple!70}
\colorlet{kit-purple50}{KITpurple!50}
\colorlet{kit-purple30}{KITpurple!30}
\colorlet{kit-purple15}{KITpurple!15}

\definecolor{KITcyanblue}{RGB}{80,170,230}
\colorlet{kit-cyanblue100}{KITcyanblue}
\colorlet{kit-cyanblue70}{KITcyanblue!70}
\colorlet{kit-cyanblue50}{KITcyanblue!50}
\colorlet{kit-cyanblue30}{KITcyanblue!30}
\colorlet{kit-cyanblue15}{KITcyanblue!15}

\title{Interference Mitigation for OFDM-based \\  Integrated Sensing and Communications with Arbitrary Modulation Formats %
\thanks{This work has received funding from the German Federal Ministry of Education and Research (BMBF) within the projects Open6GHub (grant agreement 16KISK010) and KOMSENS-6G (grant agreement 16KISK123).}
}

\author{Felix Artmann, Daniel Gil Gaviria, Benedikt Geiger and Laurent Schmalen\\
\IEEEauthorblockA{Communications Engineering Lab (CEL), Karlsruhe Institute of Technology (KIT) \\ Hertzstr. 16, 76187 Karlsruhe, Germany, Email: \texttt{daniel.gil@kit.edu}}
}

\maketitle

\begin{abstract}
    Integrated sensing and communication will be a key feature of future mobile networks, enabling highly efficient systems and numerous new applications by leveraging communication signals for sensing. In this paper, we analyze the impact of arbitrary modulation alphabets on the sensing performance of communication-centric OFDM systems as expected in the next-generation 6G networks. We evaluate existing interference mitigation techniques, such as coherent successive target cancellation, and propose an enhanced version of this algorithm. A systematic performance evaluation in multi-target scenarios, including the effects of scattering, demonstrates that our proposed interference mitigation methods achieve performance comparable to sensing-optimal constant modulus signals while utilizing higher order constellations for more efficient communications. 
\end{abstract}

\section{Introduction}

The next generation of cellular communication networks will not only use new frequency bands and larger bandwidths, but is also expected to enable advanced sensing capabilities. As a result, future cellular networks will be designed to be highly aware of their surroundings, facilitating the development of smart cities, autonomous driving, and various other applications that will shape the future of modern society. \Ac{ISAC} is undoubtedly at the core of these advancements. %

This shared use of signals and devices for both communication and sensing tasks will allow for a highly efficient utilization of physical resources, such as available bandwidth and power \cite{shatov}. %
In this work, we focus on the performance of OFDM-based ISAC systems, as \ac{OFDM} is expected to stay the main waveform in future generations of wireless communication networks. Its high bandwidth efficiency, combined with relatively low computational complexity for synchronization and equalization, makes it an attractive choice for next-generation systems \cite{wild_6g_2023, mandelli_survey_2023}. 

Every \ac{ISAC} system inherently faces a trade-off between communication and sensing performance. While communication signals are usually optimized to maximize mutual information between transmitted and received signals, sensing-optimal signals exhibit minimal randomness and typically maintain a constant modulus, making them less suitable for high-data-rate transmission.

This fundamental trade-off has been extensively analyzed in idealized scenarios, providing valuable theoretical insights \cite{liu_2023_fundamental}. However, most derivations in the literature assume idealized point targets, leading to single idealized \textit{specular} reflections. This assumption may not accurately represent real-world conditions.
In practical environments, the presence of multiple and diffuse reflections on rough surfaces (i.e., scattering) introduces further challenges that can degrade sensing performance \cite{geiger2025}.

To tackle the challenges of deploying \ac{ISAC} systems in complex scenarios, we propose an iterative method to mitigate the performance degradation and assess its effectiveness in more realistic settings. Specifically, we consider scenarios with parameters that align with the characteristics of expected near-future cellular communication networks and utilize ray tracing to simulate multiple targets with scattered reflections, moving beyond the conventional assumption of a single specular reflection. This approach provides a more comprehensive understanding of \ac{ISAC} performance in practical environments and offers valuable insights for enhancing 6G and beyond communication systems.

\vspace{-0.2cm}
\section{System model}
We consider a monostatic OFDM-ISAC setup, where the transmitter and sensing receiver are co-located. This configuration ensures that the transmitted signal is fully known to the sensing receiver and no additional synchronization is required. While we focus on a single-input single-output system, the effects observed  also apply to multiple-input multiple-ouptut systems.
We denote matrices in bold font, e.g., $\mathbf{X}$, and random variables are written in sans-serif font, e.g., $\RV{X}$. 

\subsection{OFDM Transmit Singal}
An OFDM symbol consists of $N$ orthogonal subcarriers separated by a subcarrier spacing $\Delta f$. Each subcarrier $n$ within an OFDM symbol indexed in time by $m$ is modulated by a random complex symbol $\RV{X}_{n,m}$ out of a modulation alphabet $\mathcal{X} \subset \mathbb{C}$. The alphabet is normalized such that $\mathbb{E}_\RV{X}\mleft\{ |\RV{X}|^2 \mright\} =1$ holds.
We define an OFDM frame as $M$ consecutive OFDM symbols, which we arrange into a transmit matrix 
\begin{align}
  \bm{\RV{X}}=
  \begin{pmatrix}
    \RV{X}_{0,0} & \hdots & \RV{X}_{0,M-1}\\
    \vdots  & \ddots & \vdots \\
    \RV{X}_{N-1,0} & \hdots & \RV{X}_{N-1, M-1}
  \end{pmatrix}\text{.}
\end{align}

The samples of the time domain signal $\RM{x}$ are generated by applying an \ac{IFFT} along the subcarriers
\begin{align}\label{eq:a_tilde}
\RV{x}_m[k] &= \frac{1}{N}\sum_{n=0}^{N-1} \RV{X}_{n,m} \cdot \e^{\j 2\pi nk}\text{,} 
& 0\leq  k \leq N-1 \text{,}
\end{align}
where $k$ is the discrete-time index.
A \ac{CP} is subsequently added before the physical baseband signal $s(t)$ is generated by a \ac{DAC} and upconverted to a carrier frequency~$f_\mathrm{c}$. We refer to the symbol duration (including \ac{CP}) as $T_\mathrm{S}$.

\subsection{Sensing Channel}

The sensing channel $h(t,\tau)$ can be modeled as a superposition of $L$ independent reflections 
    \begin{align}
        h(t, \tau)&=\sum_{\ell=0}^{L-1}a_\ell\delta(\tau-\tau_\ell)\cdot\e^{\j \left(2\pi f_{\mathrm{D},\ell} t +\varphi_\ell\right)}\text{,}
    \end{align}
    with amplitude $a_\ell$, time delay $\tau_\ell$, Doppler shift $f_{\mathrm{D},\ell}$ and phase shift $\varphi_\ell$, leading to the receive signal in time domain
    \vspace{-0.1cm}
    \begin{align}
        r(t) %
            &=\sum_{\ell=0}^{L-1} a_\ell s(t-\tau_\ell) \, \e^{\,\j \left(2\pi f_{\mathrm{D},\ell} t +\varphi_\ell\right) } + \RV{w}(t)\text{,}
     \end{align}
     where $\RV{w}(t)\sim\mathcal{CN} (0, \sigma_{\RV{w}}^2)$ is \ac{AWGN} with variance $\sigma_{\RV{w}}^2$.

     After sampling $r(t)$ at a sampling frequency $f_\mathrm{s}$ and removing the cyclic prefix,  the time discrete receive signal can be expressed as
    \begin{align}
        \RV{y}[k] &= \sum_{\ell=0}^{L-1} a_\ell \cdot \RV{x}\left[k-\tau_\ell f_\mathrm{s}\right] \, \e^{\,\j \left(2\pi f_{\mathrm{D}, \ell} k +\varphi_\ell\right) } + \RV{w}[k]\text{.}
     \end{align}

     The delay $\tau_\ell$ manifests in the frequency domain as a complex oscillation along the subcarriers, leading to a receive matrix $\RM{Y}$ with entries 
     \begin{align}\label{eq:Y_matrix}
        \RV{Y}_{n,m}=\RV{X}_{n,m}\cdot \underbrace{\sum_{\ell=0}^{L-1}a_\ell\e^{\j 2\pi \Delta f \tau_\ell n}\e^{\j \left(2\pi T_\mathrm{S}f_{\mathrm{D}, \ell}m-\varphi_\ell\right)}}_{H_{n,m}} + \RV{W}_{n,m} \text{,}
     \end{align}
     where $\RV{W}_{n,m}$ are independent complex normal distributed entries with variance $\sigma^2_{\RV{W}}$. We note hat the inter-carrier interference (ICI) caused by the Doppler shift  $f_{\mathrm{D}, \ell}$ is negligible if the subcarrier spacing is significantly larger than the Doppler spread of the channel, typically by a factor of 10 \cite{Nuss2018}. Modern OFDM systems generally satisfy this condition in most practical scenarios \cite{mandelli_survey_2023}. 
     Introducing a channel matrix $\mathbf{H}$ with entries $H_{n,m}$ as indicated in (\ref{eq:Y_matrix}) leads to a matrix notation
     \begin{align}
        \RM{Y}=\RM{X}\circ\mathbf{H}+\RM{W}\text{,}
     \end{align}
     where $\circ$ represents the Hadamard product.
     For clarity of notation, we introduce 
     \begin{align}
         \tilde{a}_{\ell,n,m}=a_\ell \cdot \e^{\j 2 \pi \Delta f \tau_\ell n} \cdot\e^{\j\left( 2\pi T_\mathrm{S}f_{\mathrm{D}, \ell}m+\varphi_\ell\right)}\text{,}
     \end{align}
     which encapsulates the amplitude of a reflection along with the complex harmonics related to the delay $\tau_\ell$ and the Doppler shift $f_{\mathrm{D}, \ell}$, respectively.

     We assume that all reflections have a delay shorter than the duration of the cyclic prefix, ensuring that no inter-symbol interference (ISI) arises. While this assumption may not always hold, modern algorithms for mitigating ISI have been recently proposed \cite{geiger2025longRange} and are beyond the scope of this paper.

\subsection{OFDM Radar} \label{sec:ofdm_radar}

Equation (\ref{eq:Y_matrix}) shows that the delay $\tau_\ell$ of each target manifests as an oscillation along the subcarriers (i.e., the columns of $\bm{\RV{Y}}$), while the Doppler shift $f_{\mathrm{D},\ell}$ appears as an oscillation along the consecutive symbols of the frame (i.e., the rows of $\bm{\RV{Y}}$). Thus, estimating these parameters reduces to a frequency estimation along each respective axis of $\bm{\RV{Y}}$ \cite{Braun2010}.

To eliminate the transmit symbols $\bm{\RV{X}}$, we apply a \ac{MF} $ \widehat{\bm{\RV{H}}} =\bm{\RV{Y}}  \circ  \bm{\RV{X}}^*$, which maximizes the SNR of the compensated signal and serves as an unbiased estimator for $\widehat{\RM{H}}$, provided that $\mathbb{E}\mleft\{ |\RV{X}|^2\mright\}=1$, as demonstrated in \cite{geiger2025}. The entries of $\widehat{\RM{H}}$ result in
\begin{align}\label{eq:mf_H_hat}
                      \widehat{\RV{H}}_{n,m}=H_{n,m} |\RV{X}_{n,m}|^2 +\RV{W}_{n,m}\RV{X}_{n,m}^*\text{.}
\end{align}
  After an IFFT along the subcarriers, an FFT along the consecutive symbols of the frame and normalizing by the factor $M$, we obtain the range-Doppler-matrix (RDM) $\widehat{\RM{P}}$ with entries
\begin{equation}
    \widehat{\RV{P}}_{\nu,\mu} = \frac{1}{NM}  \sum_{n=0}^{N-1} \left( \sum_{m=0}^{M-1} \widehat{\RV{H}}_{n,m} \e^{-\j\frac{2\pi m \mu}{M}} \right) \e^{\j\frac{2\pi n \nu}{N}} \text{,}
\end{equation}
where $\nu$ and $\mu$ are discrete frequency and delay indexes respectively, spanning the delay-Doppler domain. The RDM has large power at bins where the targets are present. At this point, state of the art detection algorithms such as \ac{CA-CFAR} can be applied to detect potential targets. 
The grid resolution of the periodogram is limited by the bandwidth of the signal and the total duration of used frames. We use sinc-interpolation as proposed \cite{richards2005, Braun2010} to estimate the parameters $\widehat{a}_\ell$, $\widehat{\tau}_\ell$ and $\widehat{f}_{\mathrm{D,}\ell}$ of the detected targets. %

Finally, the range $\widehat{d}_\ell$ and radial velocity $\widehat{v}_\ell$ of each target are calculated by
\begin{align}
    \widehat{d}_\ell&=\frac{c_0\cdot \widehat{\tau}_\ell}{2} \text{,} \\
    \widehat{v}_\ell&=\frac{\widehat{f}_{\mathrm{D,\ell}}\cdot c_0}{2f_\mathrm{c}}\text{,}
\end{align}
respectively, where $c_0$ is the speed of light in air.

\section{Sensing and Communications Trade-off}\label{sec:trade_off}
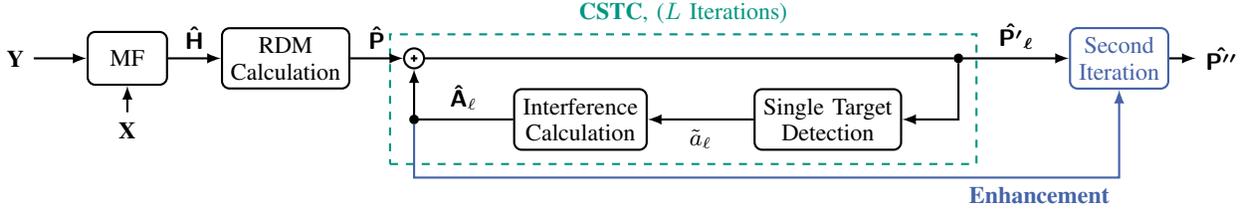
\begin{figure*}[t]
    \tikzset{%
  block/.style    = {draw, thick, rectangle, minimum height = 2em, minimum width = 3em,rounded corners=1mm, },
  sum/.style      = {draw, circle,inner sep=1pt}, %
  abzweig/.style  = {inner sep=0pt, outer sep=0pt},
  input/.style    = {coordinate}, %
  output/.style   = {coordinate} %
}
\tikzstyle{dot}=[circle,fill,draw,inner sep=1pt]
\centering
\begin{tikzpicture}[thick,node distance=0.6cm, >=latex]
  \tikzstyle{every node}=[font=\small]

    \node[block] (div) at (0,0) { %
            MF };
    \node[left=2em of div] (Y){\RM{Y}}; 
    \node[below= 1em of div] (X){\RM{X}};

    \node[block, name=RDM, right=2em of div, align=center]{RDM \\Calculation};
    \node[circle, draw, name=sum, right=2em of RDM, scale=0.8]{};
    \draw[-] (sum.west)+(1mm,0) --  ($(sum.east)+(-1mm,0)$);
    \draw[-] (sum.south)+(0,1mm) --  ($(sum.north)+(0,-1mm)$);

    \node[dot, name=dot, right=20em of sum, align=center]{};
    
    \node[block, name=det, below left=1em and 2em of dot, align=center]{Single Target\\Detection};
    \node[block, name=I_calc, left=4em of det, align=center]{Interference\\Calculation};

    \node[block, name=2Iter, right=4em of dot, align=center, KITblue]{Second \\ Iteration};
    \node[ name=out, right=1em of 2Iter, name=Hprime]{$\RM{\hat{P^{\prime\prime}}}$};

    \node[draw,dashed,KITgreen, fit= (sum) (I_calc) (det) (dot), inner sep=0.5em, align=center](CSTC) {};
    \node[above, KITgreen, align=center] at (CSTC.north){\textbf{CSTC}, ($L$ Iterations)};
    
    \draw[->] (X)--(div);
    \draw[->] (Y)--(div);
    \draw[->] (div)--node[above]{$\RM{\hat{H}}$}(RDM);
    \draw[->] (RDM)--node[above]{$\RM{\hat{P}}$}(sum);
    \draw[-] (sum)--(dot);
    \draw[->] (dot) |- (det);
    \draw[->] (det) -- node[below]{$\tilde{a}_\ell$}(I_calc);
    \draw[->] (I_calc) -- node[above]{$\RM{\hat{A}}_\ell$} (I_calc-|sum) node[dot, name=dot2]{} --(sum);
    \draw[->] (dot)--node[above]{$\RM{\hat{P^\prime}_\ell}$}(2Iter);
    \draw[->] (2Iter)--(Hprime);

    \draw[->, KITblue](dot2)--++(0,-2.2em) coordinate(x) -- (x-|2Iter) --(2Iter);
    \node[below, anchor=north east, KITblue]at(x-|2Iter) {\textbf{Enhancement}};
    
\end{tikzpicture}
    \caption{Block diagram of the CSTC algorithm and our proposed enhancement (blue)}
    \label{fig:IM_block}
\end{figure*}

In this section, we analyze the statistical properties of the estimated channel matrix $\widehat{\RM{H}}$  in a way that highlights the effect of the characteristics of the transmit alphabet $\mathcal{X}$, the presence of multiple targets and their impact on the sensing performance of the \ac{ISAC} system.

Consider the matched filter applied in (\ref{eq:mf_H_hat}). Inserting the entries of the frequency domain channel matrix $\mathbf{H}$ in (\ref{eq:Y_matrix}) yields
\begin{align}\label{eq:H_Dach}
    \widehat{\RV{H}}_{n,m} &=|\RV{X}_{n,m}|^2 \cdot \sum_{\ell=0}^{L-1} \tilde{a}_{\ell, n,m}+ \underbrace{\RV{W}_{n,m}\cdot\RV{X}_{n,m}^*}_{\coloneqq \RV{\tilde{W}}_{n,m}}\text{.}
\end{align} %

This representation provides two key insights. First, the entries of the noise matrix $\RV{W}_{n,m}$ are multiplied by the complex conjugate of the transmit symbols $\RV{X}_{n,m}^*$. As these random variables are independent, the variance of the entries $\widetilde{\RV{W}}_{n,m}$ is given by
\begin{align}\label{eq:noise_enhancement}
    \sigma^2_{\tilde{\RV{W}}}=\sigma_{\RV{W}}^2 \sigma_{\RV{X}}^2 + \underbrace{\mu_{\RV{X}}^2\sigma_{\RV{W}}^2 }_{=0} + \underbrace{\mu_{\RV{W}}^2 \sigma_{\RV{X}}^2}_{=0} \text{.} %
\end{align}
Since both the white noise and the alphabet have zero mean, the cross terms in (\ref{eq:noise_enhancement}) vanish. Furthermore, as ${\sigma_{\RV{X}}^2 =\mathbb{E}_\RV{X}\mleft\{ |\RV{X}|^2 \mright\}= 1}$, no noise enhancement occurs and ${\sigma_{\RV{\tilde{W}}}^2 = \sigma_{\RV{W}}^2}$ holds.

Second, (\ref{eq:H_Dach}) shows how the distribution of the entries in $\bm{\RV{X}}$ influences the estimate of $\widehat{\bm{\RV{H}}}$. For instance, consider a constant modulus alphabet $\mathcal{X}$, where $|\RV{X}_{n,m}|^2 = 1$ always holds. In this case, $\widehat{\bm{\RV{H}}}$ serves as an unbiased estimate of $\mathbf{H}$ in the presence of \ac{AWGN}. 
To analyze the case of a non-constant modulus alphabet, we express $|\RV{X}_{n,m}|^2 = 1+\RV{D}_{n,m}$, where $\RV{D}_{n,m}$ represents the deviation of each symbol's power from unit power. Applying this to (\ref{eq:H_Dach}) yields
\begin{align}\label{eq:H_interference}
    \RV{\widehat{H}}_{n,m} &=\sum_{\ell=0}^{L-1} \tilde{a}_{\ell,n,m}+ \RV{D}_{n,m} \cdot \sum_{\ell=0}^{L-1} \tilde{a}_{\ell,n,m} +\RV{\widetilde{W}}_{n,m}\text{.}
\end{align}

The first summand is equivalent to the physical channel, while the second one can be interpreted as an additive interference term. This interference scales with both the deviation of the transmitted symbols $\RV{D}_{n,m}$ from unit power and the power of the target reflections. For each target $\ell$, the power of this interference signal $P_{\mathrm{I,}\ell}$ is 
\begin{align}
    P_{\mathrm{I,}\ell}&=\mathrm{Var}\mleft\{\RV{D}_{n,m}  \cdot \tilde{a}_{\ell,n,m}  \mright\}\\
                    &=\mathrm{Var}\mleft\{\left(|\RV{X}_{n,m}|^2-1\right)  \cdot  \tilde{a}_{\ell,n,m} \mright\}\\
                    &=|a_\ell|^2 \cdot\left( \mathbb{E}\left\{ |\RV{X}_{n,m}|^4\right\} - \mathbb{E}\left\{ |\RV{X}_{n,m}|^2 \right\}^2\right)\\
                    &=|a_\ell|^2\cdot \left(\kappa -1 \right)\text{,}
\end{align}
where $\kappa$ denotes the \textit{kurtosis},  which  is  equivalent  to  the fourth order  moment of the input distribution  if the alphabet has  unit  power and  zero  mean. A constant modulus constellation results in $\kappa = 1$, which leads to $P_{\mathrm{I},\ell}=0$ and indicates the absence of additive interference. For other constellations, the second summand in (\ref{eq:H_interference}) behaves as additional Gaussian interference in the RDM, as demonstrated in \cite{geiger2025}.

\section{Mitigation of Modulation-dependent Additive Interference} \label{sec:Interference_Mitigation}

The derivations in Sec. \ref{sec:trade_off} demonstrate that constant modulus modulations are optimal for sensing tasks. However, incorporating the amplitude as an additional degree of freedom can significantly enhance the communication performance. To address the resulting performance degradation at the sensing receiver when using arbitrary modulation alphabets, we study interference mitigation algorithms.
In realistic scenarios with multiple targets at varying distances and radar cross sections, strong reflections (i.e., large $a_\ell$) can cause severe interference, potentially masking weaker reflections and rendering their detection unfeasible. In order to mitigate this degradation, coherent successive target cancellation (CSTC) was proposed in \cite{Braun2010}.

\subsection{Coherent Successive Target Cancellation (CSTC) }
The goal of the interference mitigation algorithm is to detect and coherently subtract the effect of each echo from the RDM $\RM{\hat{P}}$. Its basic principle is depicted in Fig.~\ref{fig:IM_block}. 

First, the matched filter is applied and an estimate of the RDM is calculated as described in Sec. \ref{sec:ofdm_radar}. Each target is detected individually yielding an estimate $\hat{\tilde{a}}_\ell$, including the amplitude $\hat{a}_\ell$, the delay $\hat{\tau}_\ell$ the Doppler shift $\hat{f}_{\mathrm{D,}\ell}$ and the phase shift $\hat{\varphi}_\ell$. 
Based on these parameters, we synthesize an RDM matrix $\RM{A}_\ell$ that contains the effect of each target $\ell$ on every bin of the estimated RDM $\hat{\RM{P}}$. The entries of  $\RM{A}_\ell$ can be reconstructed applying the model in (\ref{eq:H_Dach})
\begin{align}
        \hat{\RV{A}}_{\ell, \mu,\nu } = \mathrm{IFFT}_n\left(\mathrm{FFT}_m\left( |\RV{X}_{n,m}|^2 \cdot \hat{\tilde{a}}_{\ell,n,m} \right)\right)\text{.}
\end{align}
For each of the $L$ targets, a new estimate of the RDM $\hat{\RM{P^{\prime}}}_\ell$ is calculated by iteratively subtracting the $\ell-1$ already estimated interference matrices $\RM{\hat{A}}_{\ell^\prime}$ from the current estimate $\hat{\RM{P}}$ as in
\begin{align}
    \hat{\RM{P}^{\prime}}_\ell=\hat{\RM{P}}-\sum_{\ell^\prime=0}^{\ell-1}\hat{\RM{A}}_{\ell^\prime}\text{.}
\end{align}
This way, the interference caused by each target is removed immediately after its detection, enabling a more precise detection of the further targets.

\subsection{Enhanced CSTC (ECSTC)}
The  detection of weak targets clearly profits from the regular CSTC algorithm as described above, since the interference caused by early detected targets is removed. Nevertheless, when first estimating the strong targets, the interference cause by the weaker ones cannot be compensated. The estimation of delay and Doppler shift for the stronger targets is thus corrupted by the cumulative interference of all weaker ones. In order to avoid this degradation, we propose to enhance the CSTC algorithm by adding a second iteration of the target parameter estimation, where the interferences matrix of all previously estimated targets is removed for each range and Doppler shift estimation as in
\begin{align}
    \hat{\RM{P}}^{\prime\prime}_\ell=\hat{\RM{P}}-\smashoperator{\sum_{\forall  \ell^{\prime} \in \{1, ..., L-1 \} \backslash \{\ell\} }}\hat{\RM{A}}_{\ell^\prime} \text{.}
\end{align}

During this step, the estimated target parameters $\hat{\tilde{a}}_\ell$ from the first iteration are used.  No additional calculation of the interference matrices $\hat{\RM{A}}_\ell$ is required and the only additional computational complexity corresponds to the sum of all  $\hat{\RM{A}}_{\ell^\prime}$. 

\section{Simulation Setup}

\subsection{OFDM Parameters}

\begin{figure}[]
  \centering
  \includegraphics[width=0.8\linewidth]{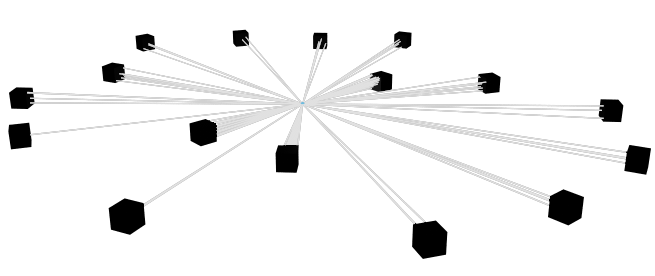}
  \caption{Simulation scene using Sionna RT \cite{hoydis2023sionna}. The distance and velocity of the cube shaped targets are uniformly distributed within $d_\ell\in [45 \text{--}145] \mathrm{m}$ and $v_\ell\in [-50 \text{--}50] \mathrm{m/s}$ respectively. The ISAC-system is located at the center, multiple rays (gray) are scattered back from each target (black cubes)}
  \label{fig:ncubes}
\end{figure}

For our simulations, we focus on the \SI{3.5}{\giga\hertz} band as it is expected to be used for 6G mobile networks~\cite{mandelli_survey_2023}. This band offers the best coverage and efficiency due to its low propagation loss. Additionally, it can be readily exploited by the existing 5G infrastructure. Tab~I lists the key parameters for a potential ISAC operation in this frequency band \cite{mandelli_survey_2023}. 

\begin{table}[h]
    
  \begin{center}
  \textsc{TABLE I} \\ \textsc{Expected parameters at $f_\mathrm{c}=\SI{3.5}{\giga\hertz}$ for 6G-OFDM systems}
    \label{tab:ofdm_parameters}

 \begin{tabular}{ p{2.4cm} c p{2cm}  }
    \toprule %
    \textbf{Parameter} & \textbf{Symbol} &\textbf{Value} \\
    \midrule %
    Subcarriers &$N$  & \num{6552} \\
    OFDM symbols &$M$  & \num{96} \\
    Subcarriers spacing &$\Delta \! f$ &  $\SI{30}{\kilo\hertz}$  \\
    Bandwidth &$B$ &  $\SI{200}{\mega\hertz}$ \\
    Cyclic prefix length & & \num{468} samples \\
    \bottomrule %
\end{tabular}

  \end{center}
\end{table}
\subsection{Figures of Merit}

In this paper,  we evaluate the \ac{MSE} between the true and estimated  target distance~$\hat{d_\ell}$ and \mbox{velocity~$\hat{v}_\ell$} 
\begin{align}
    \text{MSE}_d &= \frac{1}{K} \sum_{i=1}^K \left|d_i - \hat{d}_i \right|^2 \text{,}\\
    \text{MSE}_v &= \frac{1}{K} \sum_{i=1}^K \left|v_i - \hat{v}_i \right|^2 \text{,}
\end{align}
respectively. We calculate the \ac{MSE} by averaging over $K$ independent simulations.

\begin{figure}[t]
      \centering
      \begin{subfigure}[t]{0.9\columnwidth}
        \begin{center}
          \hspace{-1cm}
          \begin{tikzpicture}

            \definecolor{darkgray176}{RGB}{176,176,176}
            \definecolor{darkorange25512714}{RGB}{255,127,14}
            \definecolor{forestgreen4416044}{RGB}{44,160,44}
            \definecolor{steelblue31119180}{RGB}{31,119,180}
            
            \begin{axis}[
              height=0.7\columnwidth,
            width=0.8\columnwidth,
            log basis y={10},
            tick align=outside,
            tick pos=left,
            x grid style={darkgray176},
            xlabel={Target no. $\ell$},
            xmajorgrids,
            xmin=-0.75, xmax=15.75,
            xtick style={color=black},
            minor y tick num=9,
            y grid style={darkgray176},
            ylabel={MSE$_d$ (\si{\metre^2})},
            ymajorgrids,
            ymin=1e-07, ymax=3e-05,
            ymode=log,
            yminorticks=true,
            yminorgrids,
            max space between ticks=20,
            ytick style={color=black},
            label style={font=\footnotesize},
            tick label style={font=\footnotesize},  
            ytick={1e-08,1e-07,1e-06,1e-05,0.0001},
            yticklabels={
              \(\displaystyle {10^{-8}}\),
              \(\displaystyle {10^{-7}}\),
              \(\displaystyle {10^{-6}}\),
              \(\displaystyle {10^{-5}}\),
              \(\displaystyle {10^{-4}}\)
            }, 
            legend columns=-1,
            legend style={at={(0.02,0.98)}, anchor=north west, legend cell align=left, align=left, draw=white!15!black, font=\scriptsize, column sep=1em},
            ]

              \addlegendimage{KITpurple, mark=*, mark size=2, only marks} \addlegendentry{\hspace{-1em}CSTC}
              \addlegendimage{KITgreen, mark=*, mark size=2, only marks} \addlegendentry{\hspace{-1em}ECSTC}
              \addlegendimage{KITblue, thick } \addlegendentry{\hspace{-1em}CRB}

            \addplot [semithick, KITpurple, mark=*, mark size=2, mark options={solid}, only marks]
            table {%
            0 2.22407384390945e-06
            1 2.99932988494549e-06
            2 2.92294466898402e-06
            3 4.18177908452585e-06
            4 4.27194049925751e-06
            5 4.42980144878736e-06
            6 5.55076386423453e-06
            7 5.27281941897808e-06
            8 5.44051230750292e-06
            9 4.84058112023445e-06
            10 5.66640041402674e-06
            11 5.5066980573336e-06
            12 6.15099476501815e-06
            13 5.3085263207783e-06
            14 4.54403612213927e-06
            15 2.78356074333037e-06
            };
            \addplot [semithick, KITgreen, mark=*, mark size=2, mark options={solid}, only marks]
            table {%
            0 2.33415395247421e-07
            1 3.167165437915e-07
            2 3.8959343805537e-07
            3 5.50406959278021e-07
            4 5.5199285790904e-07
            5 6.87532076522113e-07
            6 7.81637795056979e-07
            7 8.84435294898415e-07
            8 1.03922151825636e-06
            9 1.02338673372776e-06
            10 1.26300237389315e-06
            11 1.3407093080643e-06
            12 1.61140295113785e-06
            13 1.64709763723241e-06
            14 1.62945971602485e-06
            15 1.55156912687407e-06
            };
            \addplot [thick, KITblue]
            table {%
            0 1.67071465995708e-07
            1 2.03790245395616e-07
            2 2.43685209775757e-07
            3 2.90425476536148e-07
            4 3.41657583735283e-07
            5 3.94929738011224e-07
            6 4.52665446364295e-07
            7 5.17856062880052e-07
            8 5.89843469079862e-07
            9 6.75152468675355e-07
            10 7.55194634639745e-07
            11 8.39536049332971e-07
            12 9.29624328002352e-07
            13 1.02399482234278e-06
            14 1.12930468858038e-06
            15 1.24396041686692e-06
            };
            \end{axis}
            \end{tikzpicture}
            
          \caption{MSE between the true target distance $d_\ell$ and the estimate~$\hat{d}_\ell$}
          \label{fig:55_cstc_it_dist}
        \end{center}
      \end{subfigure}
      \vfil
      \begin{subfigure}[t]{.9\columnwidth}
        \begin{center}
          \vspace{0.4cm}
        \hspace{-1cm}  \begin{tikzpicture}

            \definecolor{darkgray176}{RGB}{176,176,176}
            \definecolor{darkorange25512714}{RGB}{255,127,14}
            \definecolor{forestgreen4416044}{RGB}{44,160,44}
            \definecolor{steelblue31119180}{RGB}{31,119,180}
            
            \begin{axis}[
              height=0.7\columnwidth,
            width=0.8\columnwidth,
            log basis y={10},
            tick align=outside,
            tick pos=left,
            x grid style={darkgray176},
            xlabel={Target no. $\ell$},
            label style={font=\footnotesize},
            tick label style={font=\footnotesize},  
            xmajorgrids,
            xmin=-0.75, xmax=15.75,
            xtick style={color=black},
            y grid style={darkgray176},
            ylabel={MSE$_v$ (\si{\metre^2/\second^2})},
            ymajorgrids,
            ymin=1e-06, ymax=0.000212512543399736,
            ymode=log,
            ytick style={color=black},
            ytick={1e-07,1e-06,1e-05,0.0001,0.001,0.01},
            max space between ticks=20,
            yticklabels={
              \(\displaystyle {10^{-7}}\),
              \(\displaystyle {10^{-6}}\),
              \(\displaystyle {10^{-5}}\),
              \(\displaystyle {10^{-4}}\),
              \(\displaystyle {10^{-3}}\),
              \(\displaystyle {10^{-2}}\),
            },
            legend style={at={(0.02,0.98)}, anchor=north west, legend cell align=left, align=left, draw=white!15!black, font=\scriptsize, column sep=1em},
            legend columns=-1,
          ]
          \addlegendimage{KITpurple, mark=*, mark size=2, only marks} \addlegendentry{\hspace{-1em}CSTC}
          \addlegendimage{KITgreen, mark=*, mark size=2, only marks} \addlegendentry{\hspace{-1em}ECSTC}
          \addlegendimage{KITblue, thick} \addlegendentry{\hspace{-1em}CRB}
              \addplot [semithick, KITpurple, mark=*, mark size=2, mark options={solid}, only marks]
              table {%
              0 2.10702907559141e-05
              1 3.79340959049924e-05
              2 4.05884010692424e-05
              3 4.66857681156772e-05
              4 4.71580909512652e-05
              5 5.88049882774333e-05
              6 5.69347984085484e-05
              7 5.84371573766301e-05
              8 6.17710089066258e-05
              9 5.40703129670281e-05
              10 6.66829456477653e-05
              11 7.65454938239214e-05
              12 7.57269516176185e-05
              13 7.12549980565906e-05
              14 5.85741296982475e-05
              15 3.69719341964457e-05
              };
              \addplot [semithick, KITgreen, mark=*, mark size=2, mark options={solid}, only marks]
              table {%
              0 2.61020627504733e-06
              1 4.35744741800624e-06
              2 5.28967799817999e-06
              3 6.38065991216587e-06
              4 6.6709439688276e-06
              5 8.09541150248609e-06
              6 9.63159191994587e-06
              7 1.0889628124927e-05
              8 1.22023431686219e-05
              9 1.30232473493253e-05
              10 1.39315899229009e-05
              11 1.60288492415369e-05
              12 2.03466528478271e-05
              13 2.08106357677677e-05
              14 2.17516511280407e-05
              15 2.18868232867279e-05
              };
              \addplot [thick, KITblue]
              table {%
              0 2.13059661490813e-06
              1 2.59885674913722e-06
              2 3.10762151967195e-06
              3 3.70368173585603e-06
              4 4.35702462431725e-06
              5 5.03638342980053e-06
              6 5.77266418273016e-06
              7 6.60401444379636e-06
              8 7.52204148720108e-06
              9 8.60995356528164e-06
              10 9.63069978808682e-06
              11 1.07062726369332e-05
              12 1.18551329790147e-05
              13 1.30586027312588e-05
              14 1.44015779855013e-05
              15 1.58637373381548e-05
              };
              \end{axis}
          \end{tikzpicture}
  
          \caption{MSE between the true target velocities $v_\ell$ and the estimate~$\hat{v}_\ell$}
          \label{fig:55_cstc_it_velo}
        \end{center}
      \end{subfigure}%
    
  \caption{MSE per target at $\SNR_Y = \SI{0}{\decibel}$.}
  \label{fig:55_cstc_it}
\end{figure}

\begin{figure*}[b!]
  \input{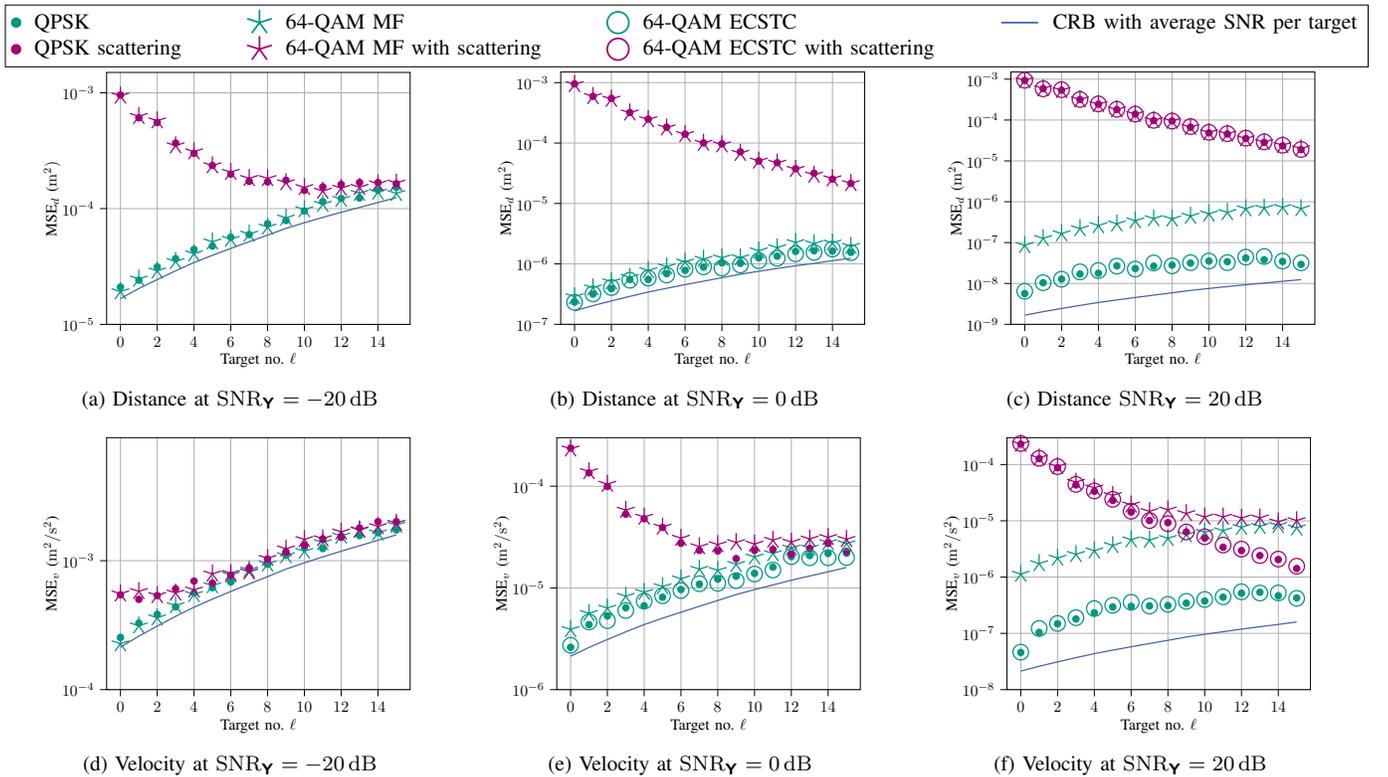}
  \caption{MSE between the true target distance $d_\ell$ and the estimate~$\hat{d}_\ell$}
  \label{fig:55_compareSNR}
\end{figure*}

\subsection{Ray Tracing Simulations}

In order to compare how the proposed interference mitigation algorithms perform in idealized simulations with only specular reflections and to simulate the effect of realistic scenarios with scattering surfaces, we build a simple scene (see Fig. \ref{fig:ncubes}) in the ray tracing simulator Sionna RT \cite{hoydis2023sionna}. 

The monostatic ISAC system is positioned at the center of the scene and is assumed to radiate isotropically. In each iteration of the simulation, 16 identical cubes with an edge length of 8 meters are placed at uniformly distributed distances $d_\ell\in [45, 145],\mathrm{m}$ with radial velocities uniformly distributed within $v_\ell\in [-50,  50],\mathrm{m/s}$. Each cube is identical in size, and its orientation is such that one face always remains orthogonal to the radial component from the origin. This ensures similar conditions for all cubes, which therefore have the same effective radar cross section.

Firstly, the cubes are assumed to cause a single mirror-like (i.e., specular) reflection and thus behave as ideal point targets like assumed in (\ref{eq:Y_matrix}). Subsequently, we use scattering in \emph{Sionna}, which accounts for diffuse reflections. In this model, a surface can reflect radiation in random directions with a given probability, which we set to $P_\mathrm{scatt}=10^{-3}$. Additionally we set the scattering coefficient to $0.9$ meaning that $90\%$ of the incident energy is reflected diffusely.  As a result, each cube generates not just a single reflection, but a set of rays spread across its area, as illustrated in Fig.~\ref{fig:ncubes}.
Each simulated point is averaged over $K=\num{1000}$ iterations of the scene.

\section{Simulation Results}

\subsection{Point Target Simulations}

First, we compare the performance of both variations of the CSTC algorithm presented in Sec. \ref{sec:Interference_Mitigation} when applied to the 16-cube scene shown in Fig. \ref{fig:ncubes}. We define the receive \ac{SNR} as
\begin{align}
    \SNR_Y=\frac{\sum_{\ell=0}^{L-1} |a_\ell|^2}{\sigma_{\RV{W}}^2}\text{,}
\end{align}

We set  up the scene such that $\SNR_Y=\SI{0}{\decibel}$ and use a \ac{64-QAM} alphabet for communications to validate the effect caused by a non-constant-modulus alphabet demonstrated in Sec. \ref{sec:trade_off}. The targets are processed from nearest to farthest, with the interference mitigation algorithms applied sequentially as illustrated in Fig.~\ref{fig:IM_block}.

The \ac{MSE} of the estimates for each target is shown in Fig.~\ref{fig:55_cstc_it}, along with the \ac{CRB} for reference. The \ac{CRB} represents the lowest theoretical variance for the estimate of distance and velocity and is given by 
\begin{align}  \label{eq:CRB}
    \text{Var}\left\{ \RV{\hat{d}} \right\} &\geq \frac{6 \sigma^2_{\RM{W}}}{(N^2-1)NM} \left(\frac{\text{c}_0}{4 \pi \Delta \! f} \right)^2 ,\\
    \text{Var}\left\{ \RV{\hat{v}} \right\}  &\geq \frac{6 \sigma^2_{\RM{W}}}{(M^2-1)NM} \left(\frac{\text{c}_0}{4 \pi T_\mathrm{S} f_\text{C}} \right)^2 ,
\end{align}
respectively \cite{Braun2010}.

The results indicate that the first iteration of the CSTC algorithm has a significant gap to the \ac{CRB}. This can be attributed to near targets experiencing substantial interference from all other targets, which can only be compensated for in later iterations. At the same time, the error of early parameter iterations adds to the error of the estimates of weaker targets. The estimate of range and velocity from far-away targets profit the most from this algorithm but suffer a higher propagation loss. This phenomenon has also been observed in \cite{Braun2010}.  

The enhanced CSTC algorithm demonstrates a clear improvement after the second iteration, yielding \ac{MSE} values very close to the \ac{CRB}. The small remaining gap to this limit may be due to residual errors after interference mitigation, caused by the imperfect estimation of target parameters through interpolation. 
\subsection{Scattering Simulations}

Finally, we introduce scattering into the simulation and evaluate the performance using both a constant modulus alphabet such as \ac{QPSK} and a non-constant amplitude alphabet (64-QAM). The evaluation is conducted at $\SNR_Y$ values of $\SI{-20}{\decibel}$, $\SI{0}{\decibel}$, and $\SI{20}{\decibel}$. The results for the \ac{MSE} of each target’s estimation are presented in Fig.~\ref{fig:55_compareSNR}, offering insights into the effectiveness of the compensation algorithms.

For idealized specular targets (green), the performance in the low SNR regime (Fig. \ref{fig:55_dist_SNR-20} and \ref{fig:55_vel_SNR-20}) is primarily limited by \ac{AWGN}, resulting in a similar performance between QPSK and 64-QAM. As expected, both modulation schemes yield \ac{MSE} values that closely approach the \ac{CRB}, since the MF compensation effectively maximizes the SNR. However, when scattering is introduced (purple), interference effects become more pronounced, especially for the nearest targets. This results from multiple small reflections occurring across a wide range of angles, which are more prominent at shorter distances. Since the \ac{MSE} is still primarily influenced by \ac{AWGN}, the range and velocity estimation errors observed with higher-order constellations remain comparable to those obtained using the constant modulus QPSK alphabet. Therefore, the use of our enhanced algorithm is not required in this regime and is omitted in Fig. \ref{fig:55_dist_SNR-20}.

In the high SNR regime (Fig.~\ref{fig:55_dist_SNR20} and \ref{fig:55_vel_SNR20}), the \ac{MSE} is predominantly influenced by interference caused by the targets within the channel. As a result, a clear performance gap emerges between QPSK (dots) and 64-QAM (stars). For specular reflections (green), the enhanced CSTC algorithm demonstrates its effectiveness by suppressing the interference and yielding an \ac{MSE} that matches the \ac{MSE} of the constant modulus signal. This result highlights the ability of the ECSTC algorithm to support high-order, non-constant-modulus constellations for improved communication performance, while preserving sensing accuracy at a level previously attainable only with constant-modulus modulation schemes.

When considering scattering (purple), the CSTC algorithm offers limited improvement, as it is designed for a single reflection per target. The remainder of the back scattered still causes noticeable interference, leading to higher MSE values. Nevertheless, the use of ECSTC always leads to a performance comparable with the case of constant-modulus alphabets as can be seen in Fig.~\ref{fig:55_vel_SNR20}.

\section{Conclusion}
This work presents a comprehensive investigation of OFDM-ISAC system performance. This includes a formal derivation for arbitrary constellations in multi-target scenarios, and a validation through ray tracing simulations based on the anticipated system setup for the next-generation 6G mobile communications. Through our analysis, we demonstrate that the reflected signals introduce additive interference that can degrade target estimation for modulation formats with non-constant modulus. The power of this interference scales with both the kurtosis of the transmitted symbol distribution and the amplitude of the reflected signals. In high-SNR scenarios, this interference becomes the dominant factor, surpassing the impact of the \ac{AWGN}.

Furthermore, we test the coherent successive target cancellation proposed by \cite{Braun2010} and introduced an enhancement to this algorithm through a second iteration. This refinement improves the estimation accuracy of each target by compensating for interference from all others, independently of their distance, magnitude or the order of the parameter estimation.

Finally, we compared the performance of these algorithms in a multi-target scenario using ray tracing simulations, both with and without scattering effects from targets with large and rough reflective areas. The enhanced CSTC algorithm effectively mitigates interference in the absence of scattering, always achieving performance comparable to that of a sensing-optimal constant modulus signal. Our findings contribute to the development of highly accurate and efficient OFDM-ISAC systems for future mobile communication networks.

\bibliographystyle{IEEEtran}
\bibliography{references_ofdm.bib}

\begin{thebibliography}{10}
\providecommand{\url}[1]{#1}
\csname url@samestyle\endcsname
\providecommand{\newblock}{\relax}
\providecommand{\bibinfo}[2]{#2}
\providecommand{\BIBentrySTDinterwordspacing}{\spaceskip=0pt\relax}
\providecommand{\BIBentryALTinterwordstretchfactor}{4}
\providecommand{\BIBentryALTinterwordspacing}{\spaceskip=\fontdimen2\font plus
\BIBentryALTinterwordstretchfactor\fontdimen3\font minus
  \fontdimen4\font\relax}
\providecommand{\BIBforeignlanguage}[2]{{%
\expandafter\ifx\csname l@#1\endcsname\relax
\typeout{** WARNING: IEEEtran.bst: No hyphenation pattern has been}%
\typeout{** loaded for the language `#1'. Using the pattern for}%
\typeout{** the default language instead.}%
\else
\language=\csname l@#1\endcsname
\fi
#2}}
\providecommand{\BIBdecl}{\relax}
\BIBdecl

\bibitem{shatov}
V.~Shatov \emph{et~al.}, ``Joint radar and communications: Architectures, use
  cases, aspects of radio access, signal processing, and hardware,'' \emph{IEEE
  Access}, vol.~12, pp. 47\,888--47\,914, 2024.

\bibitem{wild_6g_2023}
T.~Wild, A.~Grudnitsky, S.~Mandelli, M.~Henninger, J.~Guan, and F.~Schaich,
  ``{6G} {integrated} sensing and communication: From vision to realization,''
  in \emph{Proc. European Radar Conference ({EuRAD})}, Sep. 2023, pp. 355--358.

\bibitem{mandelli_survey_2023}
S.~Mandelli, M.~Henninger, M.~Bauhofer, and T.~Wild, ``Survey on integrated
  sensing and communication performance modeling and use cases feasibility,''
  in \emph{Proc. International Conference on 6G Networking (6GNet)}, May 2023.

\bibitem{liu_2023_fundamental}
Y.~Xiong, F.~Liu, Y.~Cui, W.~Yuan, T.~X. Han, and G.~Caire, ``On the
  fundamental tradeoff of integrated sensing and communications under
  {Gaussian} channels,'' \emph{{IEEE} Trans. Inf. Theory}, vol.~69, no.~9, pp.
  5723--5751, 2023.

\bibitem{geiger2025}
B.~Geiger, F.~Liu, S.~Lu, A.~Rode, and L.~Schmalen, ``Joint optimization of
  geometric and probabilistic constellation shaping for {OFDM-ISAC} systems,''
  in \emph{Proc. IEEE International Symposium on Joint Communications and
  Sensing (JC\&S)}, Oulu, Finland, Jan. 2025.

\bibitem{Nuss2018}
B.~Nuss, J.~Mayer, and T.~Zwick, ``Limitations of {MIMO} and multi-user access
  for {OFDM} radar in automotive applications,'' in \emph{Proc. IEEE MTT-S
  International Conference on Microwaves for Intelligent Mobility (ICMIM)},
  2018, pp. 1--4.

\bibitem{geiger2025longRange}
B.~Geiger, S.~Mandelli, M.~Henninger, D.~G. Gaviria, C.~Muth, and L.~Schmalen,
  ``Integrated long-range sensing and communications in multi target scenarios
  using {CP-OFDM},'' in \emph{Proc. Intl. ITG Conf. on Systems, Communications,
  and Coding (SCC)}, Karlsruhe, Germany, Mar. 2025.

\bibitem{Braun2010}
M.~Braun, C.~Sturm, and F.~K. Jondral, ``Maximum likelihood speed and distance
  estimation for {OFDM} radar,'' in \emph{Proc. IEEE Radar Conference}, 2010,
  pp. 256--261.

\bibitem{richards2005}
M.~A. Richards \emph{et~al.}, \emph{Fundamentals of {Radar Signal
  Processing}}.\hskip 1em plus 0.5em minus 0.4em\relax {Mcgraw-Hill New York},
  2005, vol.~1.

\bibitem{hoydis2023sionna}
\BIBentryALTinterwordspacing
J.~Hoydis \emph{et~al.} (2023) Sionna: An open-source library for
  next-generation physical layer research. [Online]. Available:
  \url{https://arxiv.org/abs/2203.11854}
\BIBentrySTDinterwordspacing

\end{thebibliography}

\end{document}